\documentclass[a4paper,11pt]{article}
\usepackage{pos}
\usepackage{lipsum} 

\title{The classification and categorisation of Gamma-Ray Bursts with machine learning techniques for neutrino detection}

\ShortTitle{GRB classification with machine learning}

\author*{Karlijn Kruiswijk}
\author{Gwenhaël de Wasseige} 

\affiliation{Centre for Cosmology, Particle Physics and Phenomenology - CP3, Universit{\'e} catholique de Louvain, Louvain-la-Neuve, Belgium}

\emailAdd{karlijn.kruiswijk@uclouvain.be}
\emailAdd{gwenhael.dewasseige@uclouvain.be}

\abstract{

While Gamma-Ray Burst (GRBs) are clear and distinct observed events, every individual GRB is unique. In fact, GRBs are known for their variable behaviour, and BATSE was already able to discover two categories of GRB from the T90 distribution; the short and long GRBs. These two categories match up with the expected two types of GRB progenitors. Nowadays, more features have been found to be able to further distinguish them, such as the hardness ratio or the presence of supernovae. However, that does not mean that it is by any means simple to categorise individual GRBs. Furthermore, more GRB categories have been theorised as well, such as low-luminosity or X-ray-rich GRBs. These different types of GRBs also could indicate a different neutrino spectrum, with different types of GRBs more likely to emit higher amounts of neutrinos. We present an ongoing effort to use machine learning to categorise and classify GRBs, searching for subpopulations that could yield a larger neutrino flux. We specifically use unsupervised learning, as it allows hidden patterns and correlations to come to light. With the help of features such as the T90, hardness, fluence, SNR, spectral index and even the full light curve and spectra, different structures and categories of Gamma-Ray bursts can be found.

\ConferenceLogo{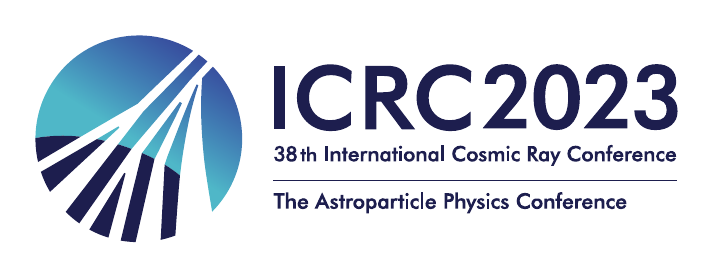}

\FullConference{The 38th International Cosmic Ray Conference (ICRC2023)\\ 26 July -- 3 August, 2023\\ Nagoya, Japan}
}

\begin{document}

\maketitle

\section{Introduction}\label{secIntro}
Gamma-Ray Burst (GRBs) have been detected by several different detectors, and none of them are the exact same. There is a great variability in the GRBs themselves, in what properties we are able to observe.  Furthermore, there are differences arising from the methods of detection; every detector is sensitive to a different energy range and they do not search the same area of sky at the same time.

Nowadays, the main detectors observing the GRBs are the Swift and Fermi telescopes. Fermi has been observing $\sim$240 GRBs per year since 2008 with its Gamma-Ray Burst Monitor (Fermi-GBM), and is able to observe GRBs in a broad energy band from 10 to 1000 keV  \cite{fermi:2020xvz}. Swift has been observing $\sim$100 GRBs with its Burst Alert Telescope (BAT), which is sensitive in the energy range from 15 to 350 keV \cite{swift:2016zny}. 

\subsection{The properties of different Gamma-Ray Bursts}\label{ssecBursts}

The observable properties of GRBs are abundant, but these properties are not available for all GRBs. This can be because of the observing constraints, the distance to the GRB or maybe even the absence of certain processes in different GRBs. 
The two main properties that are always available for GRBs are the duration of the burst and the luminosity. However, because GRBs are known to be variable in their light curves, a special measure of duration was defined: the T90, which represents the duration during which the central 90\% of the cumulative flux is detected. This T90 is able to classify what kind of GRB is observed \cite{zhang_2018}.

Generally, GRBs can be classified into two populations: the short GRBs, which have a T90 shorter than 2~s, and the long GRBs that have a T90 longer than 2~s. This split in duration also matches the two possible progenitors: merger events and massive star core collapse, respectively. This match was further confirmed with the detection of GRB 170817 the binary neutron star merger GW170817~\cite{Abbott_2017,LIGOScientific:2017ync}.  Other parameters also have an indication of what population a GRB belongs to, such that it is not solely dependent on the GRB's T90, which is also dependent on the detector observing GRBs.

In addition to the main Gamma-Ray Burst, there is also the possibility of a GRB to have a precursor emission before the main burst. Furthermore,afterglows can be observed  up to many days after the main burst, for X-ray and optical follow-up \cite{zhang_2018}. There are many properties that can be derived from the combined observations in the different energy ranges; from hardness and redshift to kilonova or supernova presence. All these parameters can help classify the GRB into one of the two populations. Especially the kilonova or supernova presence is a strong indicator of a GRB being short or long respectively.
However, the classification of the GRBs is not without debate. Many GRBs can be classified as being short or long, depending on what parameters are used or even what detector is used, as this can change how visible a GRB is. 

In addition to the two main populations, there is also the possibility of different sub-populations as well. Some of these sub-populations have been argued to consist of extended emission GRBs, low luminosity long GRBs, and choked GRBs. For example, extended emission GRBs are part of the problem with classifying GRBs as long or short just based on their T90, as they have a long T90, but portray other properties of short GRBs \cite{Zhang_2020}. These sub-populations can inhabit their own part of the parameter space of the GRBs but with not all parameters present for all GRBs and the observation-dependence of some of the parameters they can be hard to distinguish.

\subsection{Neutrino emission processes}\label{ssecNeutrinos}

While the information derived from $\gamma$-rays, X-rays and optical observations have given a better understanding on how GRBs work, there are other messengers that can give extra information for a more complete comprehension. As said before, the gravitational wave observation gave information on the progenitor that could not have been known otherwise. Another long-awaited messenger from GRBs is the neutrino. GRBs have long been expected to be neutrino emitters, as they are one of the most energetic explosions in the universe. There are different processes which can create neutrinos in GRBs.
The most known case of neutrino production is that of proton-photon interaction, where a pion is created, that can in turn decay into a muon and a neutrino. The resulting non-thermal neutrinos have energies in the order of TeV-PeV \cite{Waxman1997}.
Moreover, there is the ability to create GeV neutrinos or quasi-thermal neutrinos. These neutrinos are produced by proton-proton or proton-neutron interactions\cite{Murase_2013}, and their presence would therefore indicate some form of cosmic ray acceleration, as well as give insight into what processes are possible both inside and around a GRB.  

Unfortunately, nowadays we know that the total prompt neutrino emission from GRBs is either limited <5\% \cite{IceCube:2018omy} to even 0.1\% \cite{Abbasi:2021n1} of the observed diffuse neutrino flux, for neutrino of > TeV energies. However, here the search is done by stacking all GRBs, even though GRBs are very irregular in their $\gamma$-ray emission. It is therefore quite likely that they can differ a lot in the neutrino emission as well. For example, as the quasi-thermal neutrinos are expected to be created in a baryon-rich and dense environment, one could expect them to be more likely to be found from short, dense bursts. Being able to classify the different GRBs into different (sub-)populations, with the expected conditions needed for neutrino emission in mind, can improve the searches for neutrino emission from GRBs by giving a stronger weight to GRBs that are more likely to emit neutrinos.

\section{GRB classification}\label{secClassification}
To select the most likely neutrino-emitting GRBs, it is first necessary to classify the GRBs.
As said before, GRBs are notoriously variable in the observed phenomena. The first step in classification is trying to split the GRBs into long and short populations. Though this can be done by the simple definition of T90, as described in section\ref{secIntro}, this does not result in completely accurate populations. There are still GRBs that can belong in either category depending on what definitions are used, and even what detector observed the GRB.
This problem is exacerbated by the fact that not all observables are available for all GRBs. While some GRBs have non-detections of e.g. afterglows in different bands because they were not observed, others have no detections because no observation was made at all. 
Therefore, it is useful to make a classification scheme that relies on minimal data to make a clear classification between the two classes, without the overlap known to just relying on the T90. 

One way to do this is by looking at the prompt signal, with the use of dimension reduction algorithms like t-SNE \cite{Jespersen_2020,Jespersen_2023}. 
The t-SNE algorithm is a non-linear dimension reduction algorithm that takes into account the distances and densities of the events in multi-dimensional space. Depending on the density, it places the neighbouring events on a Student t-distribution, from which it calculates a new distance measure between each event. These events can then be fitted in a lower dimensional place with this new distance measure. In this case, the exact coordinates of each data point do not have any meaning, unlike for example with Principal Component Analysis (PCA). Instead, it is the distance and most of all the clustering of data points that are the most important. This is because these distances indicate the separation in higher dimensional space (taking into regard the density in the higher dimensional space as well). Therefore the clustering of the data points is most important, which can indicate different populations of events. 

Along with t-SNE, there are many more dimension reduction algorithms, such as PCA, UMAP (Uniform Manifold Approximation and Projection for dimension reduction) and Isometric mapping . With the exception of UMAP \cite{Jespersen_2023}, which works similarly to t-SNE, these algorithms are not able to split the data into two clear sub-populations. In the case of PCA, this can be explained by the fact that PCA is a linear reduction algorithm, creating orthogonal vectors that are in the directions that maximise the variance. These vectors do not indicate the density around these vectors, making it harder to split groups. 

\subsection{Method}\label{ssecmethod}
To make sure the methods are as similar as possible between the different detectors, similar energy bands were used for both the Swift and Fermi GRBs with data points of every 64 ms. The prompt GRB data is split into 4 energy bands: 15 to 25 keV,	25 to 50 keV, 50 to 100 keV, and 100 to 350 and either cut or padded (at the end) to 900 s.
Then, the data during the T90 (+ error) of the GRB in each energy band is concatenated to a light curve of four sequential bursts, in order of increasing energy. The new light curve can then be normalized and Fourier transformed. This is done to make sure that no direct luminosity relation, and that the placement of the burst with regards to the T90 start time is not influencing the machine learning algorithm. The resulting Fourier transform is then reduced in dimension by t-SNE to find two clusters of events. The categorisation of the events can then be calculated using the Agglomerate clustering method

\begin{figure}
\begin{minipage}[t]{.49\linewidth}
    \centering
    \includegraphics[width=\textwidth]{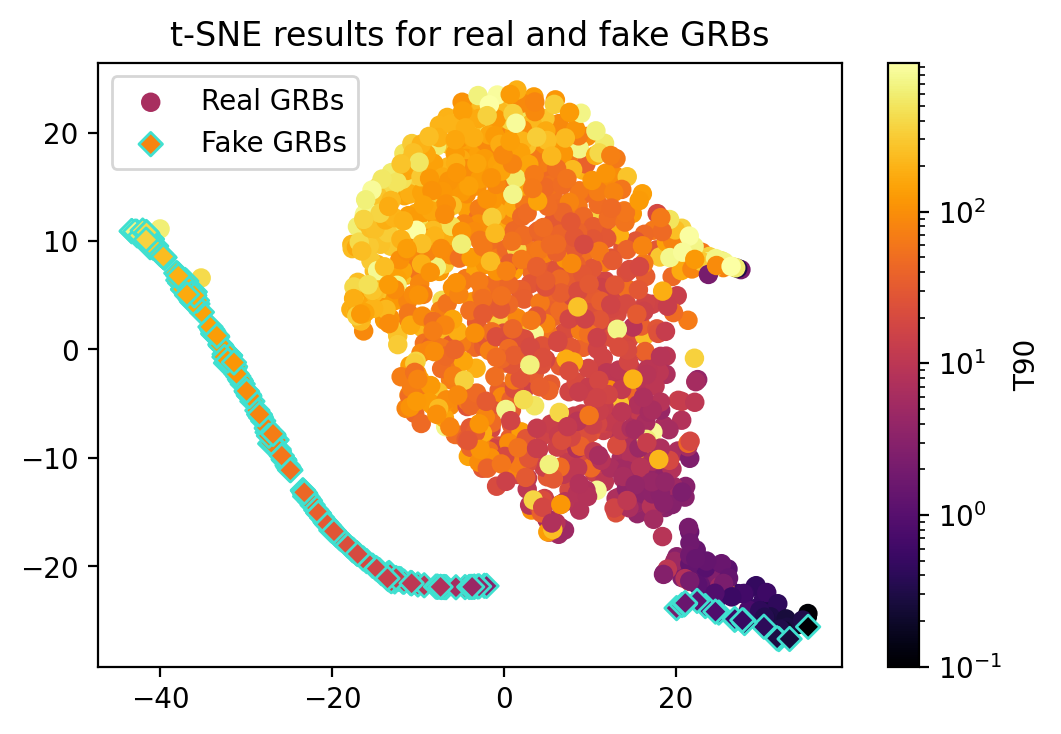}
    \caption{A t-SNE plot of Swift GRBs (circles) and of background (diamonds) without a minimum T90. The color indicates the T90. Most of the background is separated from the real GRBs, except for the short GRBs and two long GRBs with a large amount of noise.}
    \label{fGRBfake1}
\end{minipage} \hfill
\begin{minipage}[t]{.49\linewidth}
    \centering
    \includegraphics[width=\textwidth]{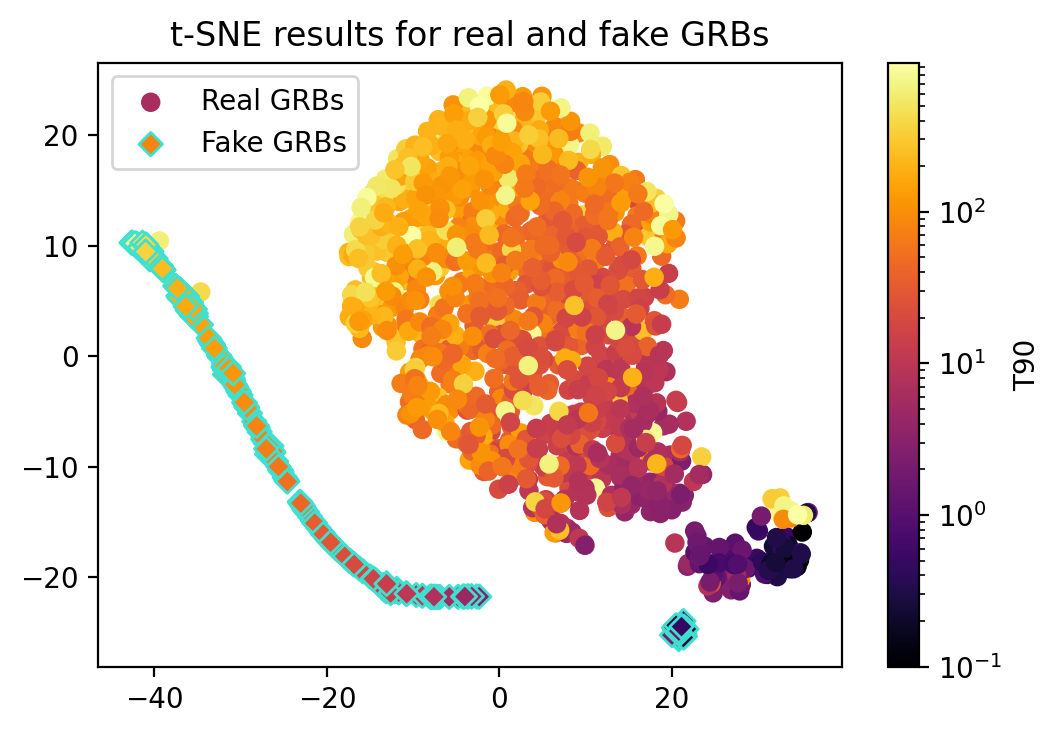}
    \caption{A t-SNE plot of the same Swift GRBs (circles) and of background (diamonds) with a minimum T90. All background is separated from the real GRBs except for the same long GRBs.}
    \label{fGRBfake2}
\end{minipage}
\end{figure}

One of the important distinctions with \cite{Jespersen_2020,Jespersen_2023}, is that, in this method, attention is paid to those events that have a T90 shorter or at the scale of the time resolution. This means that the burst can be seen as just a single jump in the light curve, which can be hard to distinguish from noise, especially when considering the input data is then the Fourier transform of a nearly empty array, which will mostly just focus on the 900 s periodicity, and might have trouble distinguishing the actual GRB. Furthermore, if one were to make fake GRBs that only consist of background data, only for fake GRBs with T90 < 0.16 s will they become hard to distinguish from the real GRBs, and start to cluster together, as can be seen in Figure \ref{fGRBfake1}. 
Therefore, in this method, in the case that T90 < 0.16 s, the T90 duration is symmetrically increased to 0.16 s. This allows for a separation between the real and fake GRBs, as well as maintaining the original separation between short and long GRBs, as seen in figure \ref{fGRBfake2}.

\subsection{Results}\label{ssecresult}
\begin{figure}[htb]
    \centering
    \includegraphics[width=0.8\textwidth]{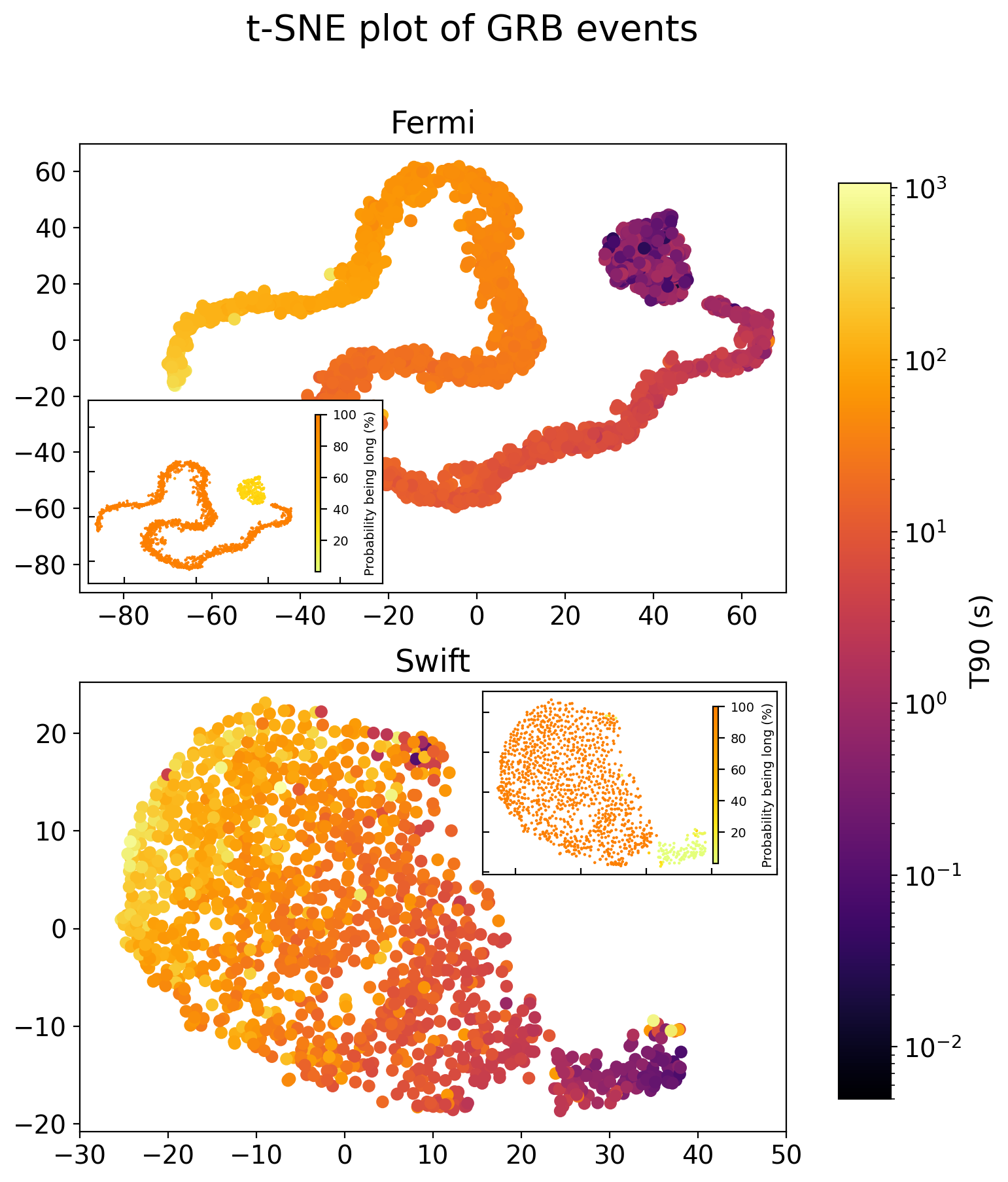}
    \caption{t-SNE plots of Fermi and Swift GRB prompt emission. The colour indicates the T90, with darker events being short GRBs. Here individual x and y values are of no particular meaning and are therefore left blank. Instead, the meaning is derived from the clustering of events. In both cases, two groups appear: one large group consisting of (mostly) long GRBs and a smaller group consisting of short GRBs. Inset shows the  probability of the GRBs falling in the long category when removing a random 20\% of the data.}
    \label{ftsne}
\end{figure}

Using the exact same methods for the GRBs of both detectors (3344 for Fermi, 1241 for Swift), a clear-cut difference between long and short GRBs can be found in both cases, as can be seen in Figure \ref{ftsne}. There is a clear split into two groups, one of which mostly consists of GRBs with a short duration and one of which mostly consists of GRBs with a long duration.

The exact split with classification on T90 or on t-SNE can be seen in the confusion matrices in tables \ref{tabconfermi} and \ref{tabconswift}. Here it is of course important to note that it is known that the T90 is not always the best measure for the classification of GRBs. Almost all GRBs that change classification between T90 and t-SNE have a T90 of ~2s, which means that their classification with just T90 is uncertain. The t-SNE method, with clear distinct clusters, is therefore able to help in these cases to give a definite classification.

\begin{table}[h]
    \begin{minipage}{.5\linewidth}
      \centering
      \caption{Fermi GRBs}\label{tabconfermi}
        \begin{tabular}{c|cc}
            & T90<2s & T90>2s\\ \hline
         Short cluster & 397 & 2 \\
         Long cluster&  33 & 2912
    \end{tabular}
    
    \end{minipage}%
    \begin{minipage}{.3\linewidth}
      \centering
        \caption{Swift GRBs}\label{tabconswift}
        
        \begin{tabular}{c|cc}
            & T90<2s & T90>2s\\ \hline
         Short cluster & 93 & 30 \\
         Long cluster&  4 & 1114
    \end{tabular}
    \end{minipage}%
    \caption*{The confusion matrices for the Fermi and Swift GRBs. Most of the GRBs match up with the expected T90, but some switch categories.}
\end{table}

\begin{table}[h]
    \centering\caption{The categories of the GRBs that are observed by both Swift and Fermi. Most are the same for both detectors.}
    \begin{tabular}{c|cc}
            & Fermi short & Fermi long\\ \hline
         Swift short& 24 & 17 \\
         Swift long &  3 & 276
\end{tabular}\label{tabmixing}
    
\end{table}

The stability of the GRB distributions is of course important here: as t-SNE is not a linear reduction algorithm, adding and removing data points can change the shapes of the clusters of events. To see that the distribution of the t-SNE algorithm is not a fluke, a random 20\% of the data can be removed before the t-SNE algorithm, to see if there are any changes in the category of different events. The amount of change from removing 20\% of the data is low, with most events staying in the same category, as can be seen in the inset of Figure \ref{ftsne}. For Swift GRBs both the long and short GRBs barely move category, remaining very stable. For Fermi GRBs the short group can sometimes get closer to the long group, which means that the clustering algorithm can confuse the clusters. However this is a limit of the clustering algorithm, and not the t-SNE per se, and even then, most of the time the category is stable. Therefore it seems that the classification with t-SNE is stable, and will hold with more GRBs added to it as well.

\subsection{Sub-populations}\label{sssubpop}

While the two main populations of GRB can come to light by the use of t-SNE, the sub-populations remain hidden. However, sub-groups of GRBs can be analyzed using graph theory, and additional properties of the GRB such as the hardness or the prompt fluence. While using the light curve as a base for the classification of the GRBs into the main population makes a lot of sense, as this classification has always been dependent on this light curve, the sub-populations do not have to differ solely in their prompt light curves. Therefore, it makes sense to use different, more global variables.

\begin{figure}[hbt]
\begin{minipage}[t]{.48\linewidth}
    \centering
    \includegraphics[width=\textwidth]{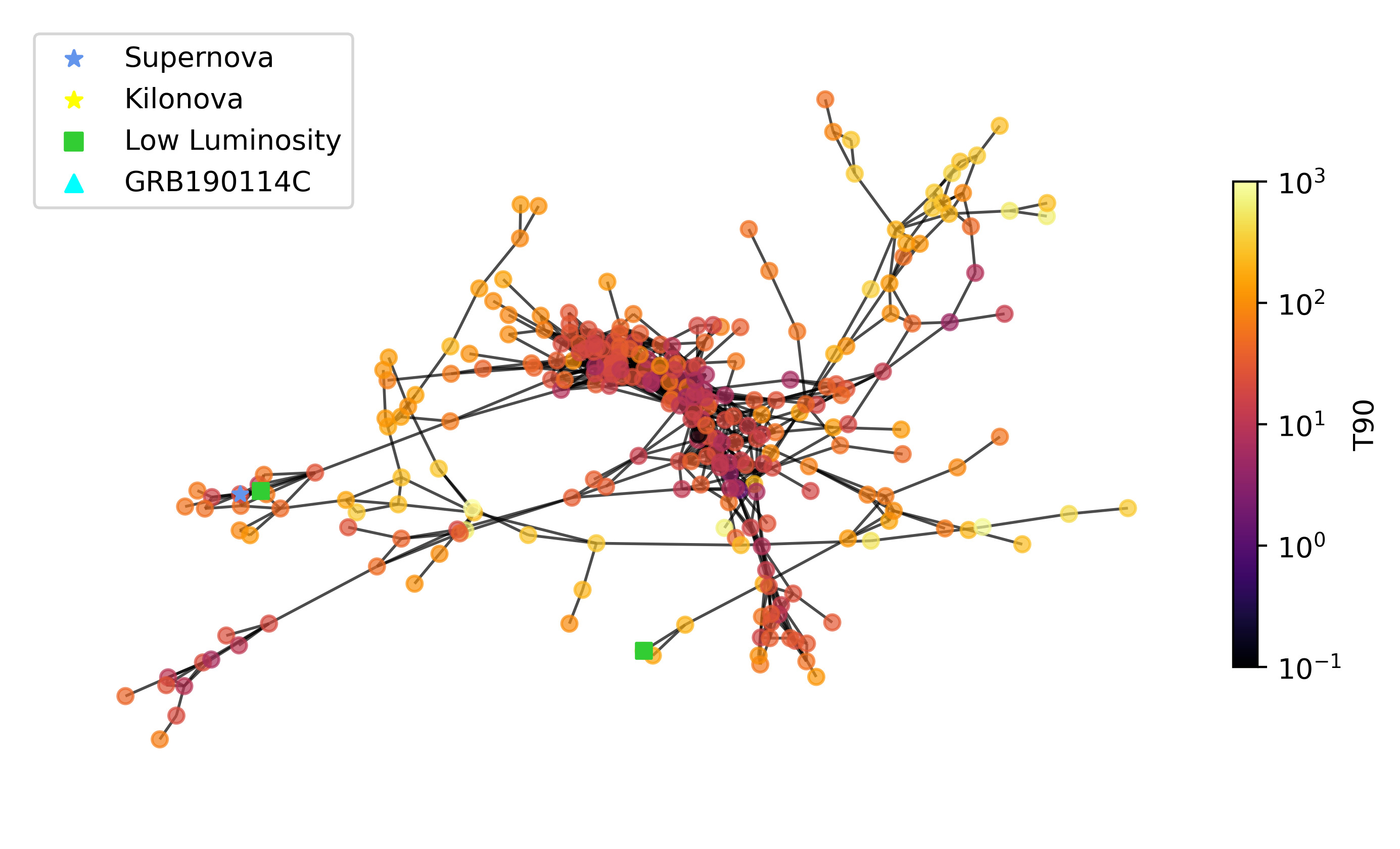}
    \caption{The largest sub-graph.}
    \label{fgraph1}
\end{minipage}\centering \quad
\begin{minipage}[t]{.48\linewidth}
    \centering
    \includegraphics[width=\textwidth]{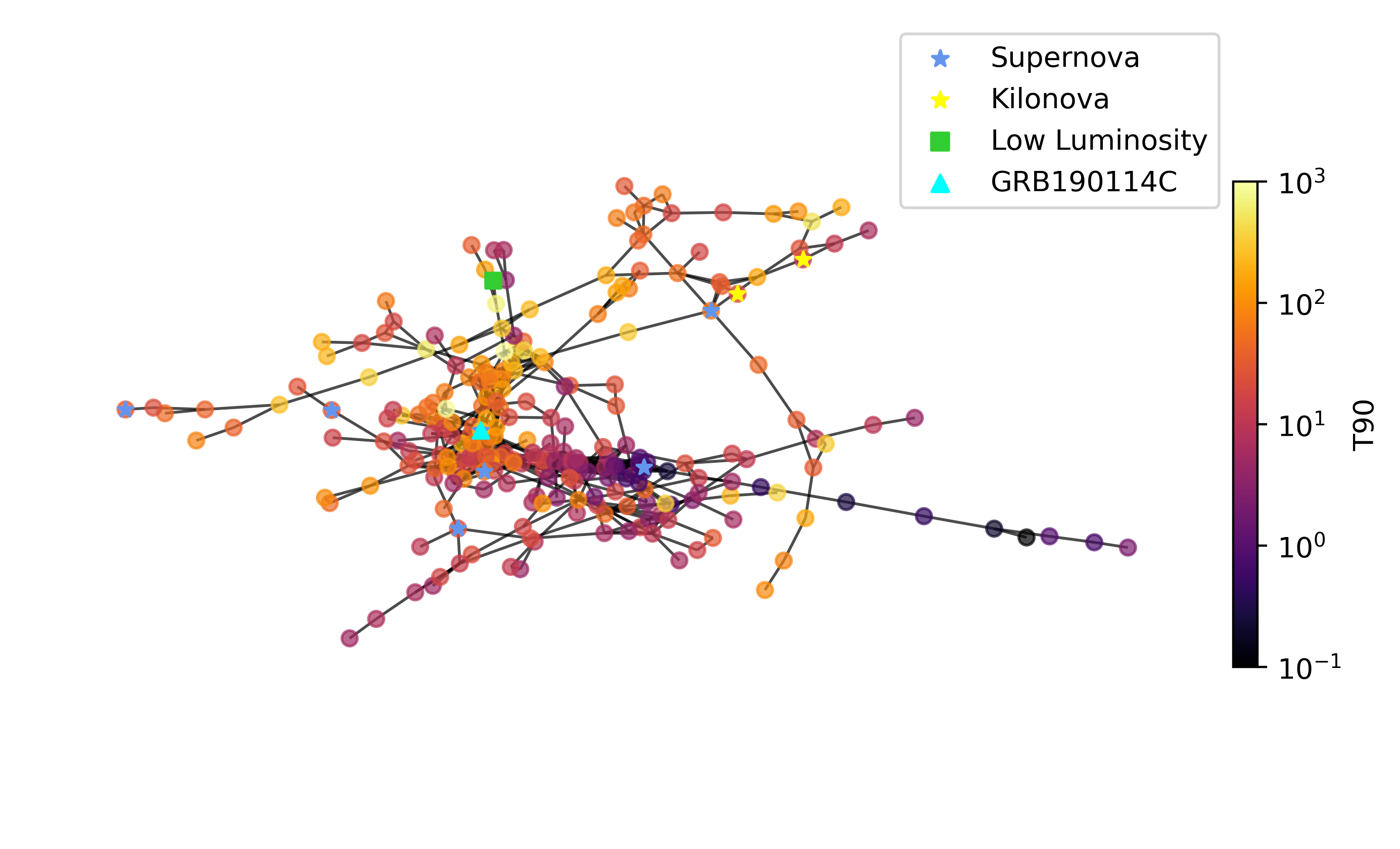}
    \caption{The second largest sub-graph.}
    \label{fgraph2}
\end{minipage}
\end{figure}

With graph theory, the similarity between individual GRBs can be visualized. GRBs are represented by individual nodes, that are connected by edges if they are similar enough. This similarity can be measured by the cosine similarity function in the parameter space: $S_c(X,Y) = \frac{X \cdot Y}{||X|| ||Y||}$. A threshold can then be placed on how large the similarity function has to be to create an edge between two GRBs, which in this case was chosen to be the value at which the second- and third-largest sub-graphs combined have the maximum number of GRBs. This process is explained more thoroughly in \cite{postergraph}.
The result of this graph theory on the Swift GRBs that were classified as long by the t-SNE method can be seen in the largest and second-largest sub-graphs in Figures \ref{fgraph1} and \ref{fgraph2}, respectively. Here the parameters used to generate the graph were: the T90, the prompt fluence, the mean to peak flux, the X-ray afterglow flux, the spectral index, and the initial X-ray temporal decay index. The supernova associated, kilonova associated, and low-luminosity GRBs are specially indicated, as well as  GRB 190114C, where TeV photons were observed, disfavouring proton synchrotron emission \cite{MAGIC:2019lau} and therefore possibly neutrino emission.

In the sub-graphs of Figures \ref{fgraph1} and \ref{fgraph2}, it is clear that there are some GRBs that are more connected than others, indicating they might be able to represent a sub-group of GRBs. Furthermore, there are some GRBs that are connected to the sub-graph through a single GRB, which could indicate further grouping on a smaller scale.

\section{Conclusions and Perspectives}\label{secCon}
We have shown that a clear-cut classification of GRBs into long and short can be made using only the prompt light curve and the t-SNE algorithm. This method is stable and can be used in the future with more GRBs added to it. Furthermore, there is a good correlation between the t-SNE classification and the standard T90 classification. With the addition of Graph theory, there is also the possibility to look for further sub-populations, as well as GRBs that can best represent these sub-populations. Further study is ongoing to look into these sub-populations and representative GRBs. We have shown that machine learning can be a valuable tool to classify GRBs and to use for population studies. Utilizing these classes for neutrino searches can allow more sensitivity for the populations of GRBs that are more likely to emit neutrinos.

\subsection*{Acknowledgments}
K.K. is a Research Fellow of the Fonds de la Recherche Scientifique - FNRS.
\bibliographystyle{ICRC}
\bibliography{references}

\providecommand{\href}[2]{#2}\begingroup\raggedright\begin{thebibliography}{10}

\bibitem{fermi:2020xvz}
A.~von Kienlin {\em et~al.}
  \href{http://dx.doi.org/10.3847/1538-4357/ab7a18}{{\em Astrophys. J.}
  {\bfseries 893} (2020) 46}.

\bibitem{swift:2016zny}
A.~Lien {\em et~al.} \href{http://dx.doi.org/10.3847/0004-637X/829/1/7}{{\em
  Astrophys. J.} {\bfseries 829} no.~1, (2016) 7}.

\bibitem{zhang_2018}
B.~Zhang, \href{http://dx.doi.org/10.1017/9781139226530}{{\em The Physics of
  Gamma-Ray Bursts}}.
\newblock Cambridge University Press, 2018.

\bibitem{Abbott_2017}
B.~P. Abbott {\em et~al.}
  \href{http://dx.doi.org/10.3847/2041-8213/aa920c}{{\em The Astrophysical
  Journal Letters} {\bfseries 848} no.~2, (Oct, 2017) L13}.

\bibitem{LIGOScientific:2017ync}
{\bfseries LIGO Scientific, and other} Collaboration, B.~P. Abbott {\em et~al.}
  \href{http://dx.doi.org/10.3847/2041-8213/aa91c9}{{\em Astrophys. J. Lett.}
  {\bfseries 848} no.~2, (2017) L12}.

\bibitem{Zhang_2020}
X.~L. Zhang {\em et~al.}
  \href{http://dx.doi.org/10.1088/1674-4527/20/12/201}{{\em Res. Astron.
  Astrophys.} {\bfseries 20} (2020) 201--212}.

\bibitem{Waxman1997}
E.~Waxman and J.~Bahcall
  \href{http://dx.doi.org/10.1103/PhysRevLett.78.2292}{{\em Phys. Rev. Lett.}
  {\bfseries 78} (Mar, 1997) 2292--2295}.

\bibitem{Murase_2013}
K.~Murase, K.~Kashiyama, and P.~M{\'{e}}sz{\'{a}}ros
  \href{http://dx.doi.org/10.1103/physrevlett.111.131102}{{\em Physical Review
  Letters} {\bfseries 111} no.~13, (Sep, 2013) }.

\bibitem{IceCube:2018omy}
{\bfseries IceCube} Collaboration, M.~G. Aartsen {\em et~al.}
  \href{http://dx.doi.org/10.1103/PhysRevLett.122.051102}{{\em Phys. Rev.
  Lett.} {\bfseries 122} no.~5, (2019) 051102}.

\bibitem{Abbasi:2021n1}
{\bfseries IceCube} Collaboration, R.~Abbasi {\em et~al.}
  \href{http://dx.doi.org/10.22323/1.395.1129}{{\em PoS} {\bfseries ICRC2021}
  (2021) 1129}.

\bibitem{Jespersen_2020}
C.~K. Jespersen {\em et~al.}
  \href{http://dx.doi.org/10.3847/2041-8213/ab964d}{{\em The Astrophysical
  Journal Letters} {\bfseries 896} no.~2, (Jun, 2020) L20}.

\bibitem{Jespersen_2023}
C.~L. Steinhardt {\em et~al.}
  \href{http://dx.doi.org/10.3847/1538-4357/acb999}{{\em The Astrophysical
  Journal} {\bfseries 945} no.~1, (Mar, 2023) 67}.

\bibitem{postergraph}
J.~Mauro and G.~de~Wasseige {\em PoS} {\bfseries ICRC2023} (these proceedings)
  1292.

\bibitem{MAGIC:2019lau}
{\bfseries MAGIC} Collaboration, V.~A. Acciari {\em et~al.}
  \href{http://dx.doi.org/10.1038/s41586-019-1750-x}{{\em Nature} {\bfseries
  575} no.~7783, (2019) 455--458}.

\end{thebibliography}\endgroup

\end{document}